\documentclass[apj]{emulateapj}

\usepackage{hyperref} 
\usepackage{amstext}
\usepackage{graphicx}

\newcommand{\be}{\begin{equation}}
\newcommand{\ee}{\end{equation}}
\newcommand{\ba}{\begin{eqnarray}}
\newcommand{\ea}{\end{eqnarray}}

\newcommand{\simgreater}{\buildrel > \over \sim}
\newcommand{\simless}{\buildrel < \over \sim}

\begin{document}

\title{Hypercritical Accretion onto a Newborn Neutron Star \\ and Magnetic Field Submergence}

\author{Cristian G. Bernal, Dany Page, and William H. Lee}
\affil{Departamento de Astrof\'{\i}sica Te\'orica, Instituto de Astronom\'{\i}a, \\
       Universidad Nacional Aut\'onoma de M\'exico,
        M\'exico, D.F. 04510, M\'exico}
\email{bernalcg@astro.unam.mx, page@astro.unam.mx, wlee@astro.unam.mx}
\begin{abstract}
We present magnetohydrodynamic numerical simulations of the late post-supernova
hypercritical accretion to understand its effect on the magnetic field of the new-born neutron star.
We consider as an example the case of a magnetic field loop protruding from the star's surface.
The accreting matter is assumed to be non magnetized and, due to the high accretion rate, 
matter pressure dominates over magnetic pressure.
We find that an accretion envelope develops very rapidly and once it becomes convectively stable
the magnetic field is easily buried and pushed into the newly forming neutron star crust.
However, for low enough accretion rates the accretion envelope remains convective for an extended period of time
and only partial submergence of the magnetic field occurs due to a residual field that is maintained at the
interface between the forming crust and the convective envelope.
In this latter case, the outcome should be a weakly magnetized neutron star with a likely complicated field geometry.
In our simulations we find the transition from total to partial submergence to occur around $\dot M \sim 10 M_\odot$ yr$^{-1}$.
Back-diffusion of the submerged magnetic field toward the surface, and the resulting growth of the
dipolar component, may result in a delayed switch-on of a pulsar on time-scales of centuries to millenia.
\end{abstract}

\section{Introduction}

Most observed neutron stars show clear evidence for the presence of strong magnetic fields.
In the case of magnetars \citep{Woods:2006fk} the estimated strength of the surface 
magnetic field is of the order of $10^{15}$ G while for the grind of the mill radio pulsar
$10^{12}$ G is a typical value.
Lower magnetic fields are, however, found in millisecond pulsars \citep{Phinney:1994uq}
and in neutron stars in low-mass X-ray binaries \citep{Psaltis:2006kx}, but in both cases
past or present, respectively, accretion is thought to be the cause of the magnetic field reduction.
There is, however, a small group of neutron stars, found in young supernova remnants
and dubbed CCOs (``Central Compact Object'', \citealt{Pavlov:2002vn}), which exhibit
little or no evidence for the presence of a magnetic field.

The case of the supernova SN1987A is also intriguing since no evidence for the presence of a
compact object has yet been found,
despite extensive searches (see, e.g., \citealt{Manchester:2007tg}).
The presence of a young energetic pulsar is clearly excluded,
but a slowly rotating neutron star, with period $P>100$ ms, and with a not too strong dipolar magnetic field,
$B_d < 10^{12}$ G, is still compatible with all current data 
\citep{Manchester:2007tg,Ng:2009kl,Larsson:2011hc}.
Such a period and low magnetic field are within the range of values inferred, when possible,
in some of the CCOs  \citep{Gotthelf:2008fu}.
The possibility that this supernova may have produced a black-hole,
as proposed, e.g., by \cite{Brown:1994fv} on the basis of a very soft dense matter 
equation of state and, consequently, a low neutron star maximum mass around
$1.6 M_\odot$, is now very slim in view of the existence of a $2 M_\odot$ pulsar
\citep{Demorest:2010dz}.

The origin of neutron star magnetic fields is still an unsolved problem
(for recent reviews, see \citealt{Spruit:2008kl,Spruit:2009qa}).
Two main mechanisms, a fossil field from the progenitor compressed
during the core-collapse and a proto-neutron star dynamo, are still
competing and are likely both needed to explain the large variety of
observed field strengths.
In these scenarios, the magnetic field generation and/or adjustment process
terminates within a minute after the neutron star's birth.

After this early field development the story is not necessarily over.
The supernova shock is still pushing its way through the outer layers of the
progenitor and, if it encounters a density discontinuity, a reverse shock
can be generated. 
Depending on its strength and on how far out it was generated, this reverse
shock can induce strong accretion onto the new-born neutron star on time scales of hours.
Notice that this delayed accretion has to be distinguished from the initial
fall-back which occurs seconds after the initial core collapse. 
A particularly favorable configuration for such late accretion is present in the
cases the progenitor had a tenuous H/He envelope surrounding a dense He core,
as in Type IIb supernovae or in SN 1987A \citep{Smartt:2009mi}.

\cite{Chevalier:1989kx} argued in favor of such late accretion in the case of SN 1987A
and developed a simple analytical model for it 
(see also \citealt{Brown:1994ly} for a similar model 
and Figure 1 of \citealt{Bernal:2010vn} for a depiction of these scenarios). 
Highly super-Eddington accretion results, in which the photons are trapped within the accretion flow
and the energy liberated by the accretion is lost through neutrino emission close to the neutron star surface.
Such a regime has been termed ``hypercritical accretion'' \citep{Blondin:1986ys,Chevalier:1989kx,Houck1991}
and requires an accretion rate $\dot{m}$ higher than about $10^3 \, \dot{m}_\mathrm{Edd}$,
where $\dot{m}_\mathrm{Edd} \sim 10^5$ g~cm$^{-2}$~s$^{-1}$ is the Eddington rate. 
We note that this regime is relevant not only for SNe, but for Gamma Ray Bursts as well 
\citep{Narayan05,Nakar07,LeeRR07,Gehrels09}, allowing rapid mass accretion onto newborn black holes 
to produce the required power to account for isotropic equivalent  luminosities that can reach 
above $10^{52}$~erg~s$^{-1}$ in the prompt emission \citep{npk01,km02,lrrp05}, 
and possibly extended emission episodes as well \citep{rosswog07, lrrlc09,metzger10}. 
In the case of core collapse events, after the reverse shock hits the neutron star surface 
a third shock develops and starts moving outward against the infalling matter.
Once this accretion shock stabilizes it will separate the
infalling matter from an extended envelope in quasi-hydrostatic equilibrium. 

Following the suggestion of \cite{Muslimov:1995nx},
\cite{Geppert:1999cr} presented simple 1D ideal MHD simulations of the effect of this post-supernova 
hypercritical accretion on the new-born neutron star magnetic field.
The result was a rapid submergence of the field into the neutron star.
After accretion stopped, the field could diffuse back to the surface and result in a delayed
switch-on of a pulsar \citep{Michel:1994bh,Muslimov:1995nx}.
Depending on the amount of accreted matter, the submergence could be so deep
that the neutron star may appear and remain unmagnetized for more than a Hubble time
\citep{Geppert:1999cr}.
This scenario was recently revisited by \cite{Ho:2011pi} and applied to study the field evolution
of the CCOs.

The 1D MHD simulations of \cite{Geppert:1999cr} were, however, carried out with some considerable simplifications.
In a previous work (\citealt{Bernal:2010vn}, Paper I hereafter) we presented results
of 2D MHD simulations of hypercritical accretion onto a magnetized neutron star surface,
with simple magnetic field configurations. 
We simulated an hypercritical accretion flow in a rectangular domain,
with the neutron star surface lying at its base, in which an initial
uniform, horizontal or vertical, magnetic field is present, as well as an horizontal field
with strength decreasing with height.
For highly-hypercritical accretion rates, $\dot{m} \simgreater 10^{10} \, \dot{m}_\mathrm{Edd}$,
complete submergence of the magnetic field was observed in a time scale of a few tens to a few
hundreds of milliseconds. 

In the present work we extend the scenarios considered in Paper I, to include an initial magnetic field
configuration which consists of a magnetic loop protruding from the neutron star surface.
We mostly consider the 2D case in a cylindrical box and one example of a 3D model.
We also extend our simulations to lower accretion rates, down to $10^{6} \, \dot{m}_\mathrm{Edd}$.
With the field strengths we employ, matter and ram pressures still strongly dominate over the 
magnetic pressure.
Moreover, we assume the accreting matter is unmagnetized.

In \S\ref{Sec:Model} we describe the numerical method and a simple analytical model.
In \S\ref{Sec:Results} we describe our results, of which an interpretation is proposed in~\S\ref{Sec:Discussion}.
Finally, we conclude in~\S\ref{Sec:Conclusions}.

\section{The Numerical Hypercritical Accretion Model }
\label{Sec:Model}

The present work is a direct extension of our previous work presented in Paper I. 
As described in Paper I, we use a customized version of the hydrodynamic code AMR FLASH2.5
\citep{Fryxell:2000uq}, with the Split 8 wave solver which solve the whole set of MHD equations.
We work in the ideal MHD regime with only numerical resistivity and viscosity.
The matter equation of state is an adaptation of FLASH's HELMHOLTZ package which includes
contributions from the nuclei, e$^- - e^+$ pairs, and radiation, plus the Coulomb correction. 
Neutrino energy losses are dominated by the e$^- - e^+$ annihilation process, but we also include
the photo-neutrino, plasmon decay, e-ion bremsstrahlung and synchrotron, processes, which are
implemented in a customized module (see Paper I).
No nuclear reactions are taken into account.

\begin{figure}
\begin{center}
\includegraphics[width=0.4\textwidth]{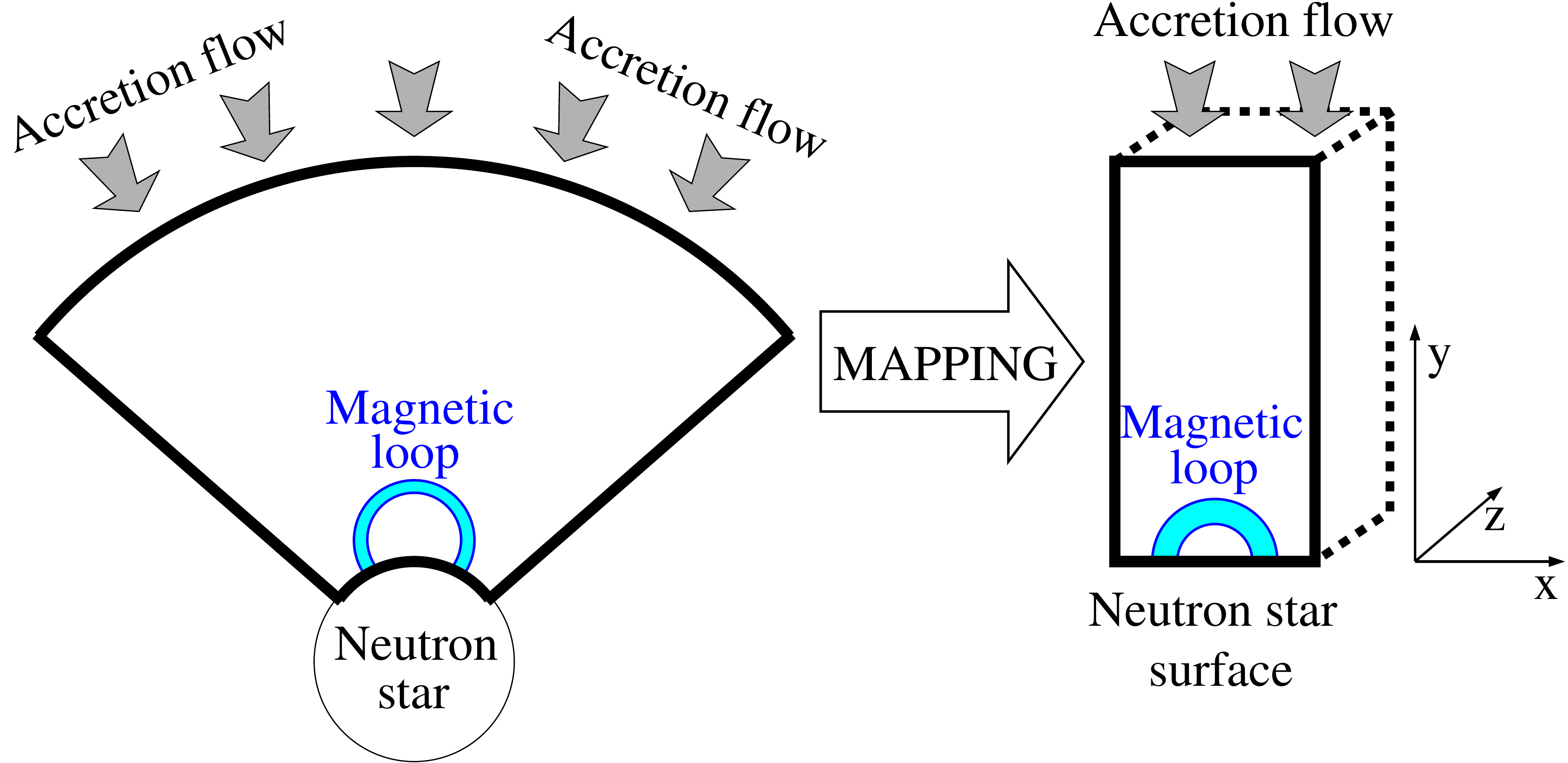}
\end{center}
\caption{Mapping of a portion of a spherically symmetric accretion flow onto a neutron star
into a 2D, or 3D, cartesian domain. Illustrated is a magnetic field loop that will react, anisotropically, to
the accretion.}
\label{Fig:1}
\end{figure}

Since the MHD equations can be solved by FLASH only in cartesian coordinates, we consider wide
columns in 2D and 3D, with a base of 
$\Delta x = 2 \cdot 10^6$ cm centered on $x=0$ in 2D, or 
$\Delta x \times \Delta z = (2 \cdot 10^6) \times (2 \cdot 10^6)$ cm$^2$ centered on $(x,z)=(0,0)$
in 3D, and a height $\Delta y$ of several times $10^6$ cm, with $y=10^6$ being the neutron star surface.
The mapping of the surrounding cone above the neutron star into a column is illustrated in Figure~\ref{Fig:1}.
We consider as magnetic initial condition a magnetic field loop, in the shape of an hemi-torus.
On the central hemi-circle of the loop the field has a strength $B_0 = 10^{12}$ G and about
it is shaped as a Gaussian, i.e., with strength $B(d) = B_0 \exp\left[-(d/R_L)^2\right)]$,
$d$ being the distance to the loop central hemi-circle and $R_L = 1$ km.
The two feet of the loop are centered at $x=-5$ and $x=+5$ km, and $z=0$ in the 3D model.
The gravitational acceleration is taken as $g_y=-GM/y^2$ and we
assume a neutron star mass of $1.4 M_\odot$.

As boundary conditions, we impose mass inflow along the top edge of the computational domain, 
and periodic conditions along the sides. 
At the bottom, on the neutron star surface, we use a custom boundary condition which enforces 
hydrostatic equilibrium (see Paper I). 
For the magnetic field, the two lateral sides are also treated as periodic boundaries, 
while at the bottom the field is frozen from the initial condition, i.e., the two feet of the loop are 
anchored into the neutron star and no field can be pushed into the star by the accretion.
On the top boundary the magnetic field is set to zero, i.e., we assume the accreting matter
to be non-magnetized.

We are interested in following the evolution of the magnetic field under the heavy hypercritical accretion
from a reverse shock (see Figure 1 in Paper I) which reaches the neutron star surface hours after the
core collapse.
The various accretion rates, $\dot{m}$, we consider will be expressed in terms of a fiducial rate
\be
\dot{m}_0 = 1.75 \times 10^{15} \; \mathrm{g \, cm^{-2} \, s^{-1}}
\label{Eq:mdot0}
\ee
and always assumed to be constant during the entire simulations.
This corresponds to a total accretion rate onto the neutron star of
\be
\dot{M}_0 \sim 350 \; M_\odot \, \mathrm{yr}^{-1}
\label{Eq:Mdot0}
\ee
and is the accretion rate originally estimated by \cite{Chevalier:1989kx} for SN 1987A.
We consider two different scenarios regarding the initial conditions for the accreting matter.
In the first case, corresponding to free fall, we start the simulation just before the reverse shock reaches the magnetic loop,
which is initially immersed in a low density, $10^2$ g cm$^{-2}$, medium, and above it the falling-back
matter has a free-fall velocity and a density obtained by mass conservation:
\be
v_\mathrm{ff} = \sqrt{2GM/y}  
\;\;\;\; \mathrm{and} \;\;\;\;
\rho_\mathrm{ff} = \dot{m}/v_\mathrm{ff}
\label{Eq:free-fall}
\ee

As we will see, after a few tens of milliseconds an accretion envelope develops which is in a
quasi-hydrostatic equilibrium, separated from the continuous accretion inflow by an accretion shock.

In the second set of scenarios, we start with this quasi-hydrostatic equilibrium envelope (QHEE hereafter)
and the magnetic loop immersed within it.
The structure of the QHEE can easily be obtained analytically,
\be
P = P_\mathrm{sh} (y_\mathrm{sh}/y)^4
,
\rho = \rho_\mathrm{sh} (y_\mathrm{sh}/y)^3
,
v = v_\mathrm{sh} (y/y_\mathrm{sh})^3
,
\label{Eq:QHEE}
\ee
where the first two values come from imposing hydrostatic equilibrium for a polytropic equation of state,
$P \propto \rho^\gamma$ with $\gamma = 4/3$, and the velocity is fixed by mass conservation.
Once the shock position $y_\mathrm{sh}$ is known (see below for its determination), $P_\mathrm{sh}$ and
$\rho_\mathrm{sh}$ are determined by the strong shock condition and $v_\mathrm{sh}$ by mass conservation
as
\be
P_\mathrm{sh} = \frac{7^2}{8} \rho_\mathrm{sh} v_\mathrm{sh}^2
, \;
\rho_\mathrm{sh} = 7 \rho_\mathrm{ff}(y_\mathrm{sh})
, \;
v_\mathrm{sh} = \frac{1}{7} v_\mathrm{ff}(y_\mathrm{sh})
.
\label{Eq:shock}
\ee
For given $M$ and $R$, and a fixed $\dot{m}$, 
the vertical location of the accretion shock, $y_\mathrm{sh}$, is controlled by energy balance
\be
\frac{GM\dot{m}}{R} = \int_R^\infty \epsilon_\nu(y) dy
\label{Eq:Ebalance}
\ee
between the accretion power and the integrated neutrino losses, per unit neutron star surface area.
Neutrino losses are dominated by e$^- - e^+$ pair annihilation for which a simple rate is
\citep{Dicus:1972ve}
\be
\epsilon_\nu = 1.83 \times 10^{-34} P^{2.25} \; \mathrm{erg \, cm^{-3} \, s^{-1}} \; .
\label{Eq:nu}
\ee
Due to the resulting strong $y$ dependence of $\epsilon_\nu$, $y^{-9}$, the value of the upper limit
in the integral of Equation~(\ref{Eq:Ebalance}) is not important and this fixes the height of the accretion shock as
\be
y_\mathrm{sh} \simeq 8 \times 10^6 \, (\dot{m}_0/\dot{m})^{10/(7 \cdot 9)} \; \mathrm{cm} \, .
\label{eq:ysh}
\ee
A stationary envelope, in quasi-hydrostatic equilibrium,
will expand, or shrink, so that the physical conditions at
its base allow neutrinos to carry away all the energy injected
by the accretion. 
Once emitted, neutrinos will act as an energy sink provided the material is optically thin to them.
Under the present conditions of density and temperature at the base of the flow, 
the fluid consists mainly of free neutrons, protons and electrons. 
The main sources of neutrino opacity are then coherent scattering off neutrons and protons and pair annihilation. 
For example, the corresponding cross section for coherent scattering \citep{Tubbs:1975fk,Shapiro:1983uq} is 
$\sigma_{\rm N}=(1/4)\sigma_0[E_{\nu}/(m_e c^2)]^2$, where $\sigma_0=1.76 \times 10^{-44}$~cm$^{2}$. 
As these are thermal neutrinos, their energy is $E_\nu \sim k_B T$, and with temperatures 
$T \simless 10^{11}$ K $\sim 10$ MeV as we will find, 
we have $\sigma_{\rm N} \simless 7 \times 10^{-42}$ cm$^2$.
The maximum densities reached at the bottom of the envelope will be below $10^{11}$ g cm$^{-3}$,
and in such conditions, the neutrino mean free-path is 
$l_{\nu}=(n_{\rm N} \sigma_{\rm N})^{-1} \simgreater 2.5 \times 10^{6}$~cm,
which is safely larger than the depth of the dense envelope, of the order of a few km.
Above this dense region the envelope density decreases rapidly, Equation~(\ref{Eq:QHEE}),
and the whole envelope is practically transparent to neutrinos.
We will, hence, ignore neutrino absorption and heating.

When using a QHEE as initial condition, the $P$, $\rho$, and $v$ profile of Equation~(\ref{Eq:QHEE})
are modified in the first two kilometers just above the neutron star surface
to smoothly turn into raising $P$ and $\rho$, and decreasing $v$ power-law profiles.

Besides been used as an initial condition in the QHEE scenarios, this analytical envelope model
will be useful in the free-fall scenarios for comparison with the numerical results when the system 
as reached a stationary state.

In Paper I we performed detailed comparisons of the hydro solver and the MHD solver with zero magnetic field
and obtained excellent agreement, as well as agreement with the above described analytical envelope
models when a stationary states had been reached. More details can be found in Paper I.

\section{Results}
\label{Sec:Results}

We will here present results from two series of simulations,
with different accretion rates.
In the first series, we consider highly hypercritical accretion rates,
$\dot{m} = (1, 10, 100) \cdot \dot{m}_0$, while in the second series
weakly hypercritical rates, 
$\dot{m} = (10^{-4}, 10^{-3}, 10^{-2}, 10^{-1}) \cdot \dot{m}_0$
are studied.
We will describe the scenario of matter initially in free-fall, with the resulting
formation of an accretion shock and the development of a QHEE, and the simplified
scenario where we start with a previously formed QHEE and follow its evolution.
With the high accretion rates of the first series we are able follow the complete evolution
of the accretion shock and the development of the QHEE in the free-fall scenario and 
compare the results with the simplified purely QHEE scenario.
In contradistinction, in the presence of the lower accretion rates of the second series,
the accretion shock would move outward to very large distances and leave the computational domain
and we are, hence, forced to restrict ourselves to the initial QHEE scenario.
Our distinction between highly and weakly hypercritical rates is thus purely numerical
and does not implies different physical conditions and/or results.
At the highest accretion rate, $100 \, \dot{m}_0$, since the accretion shock expands to small heights
we will present results of a 3D MHD simulation
while for the other cases only 2D simulations have been performed.

\subsection{Highly hypercritical accretion rates}
\label{Sec:highrates}

\begin{figure}[b]
\begin{center}
\includegraphics[width=0.45\textwidth]{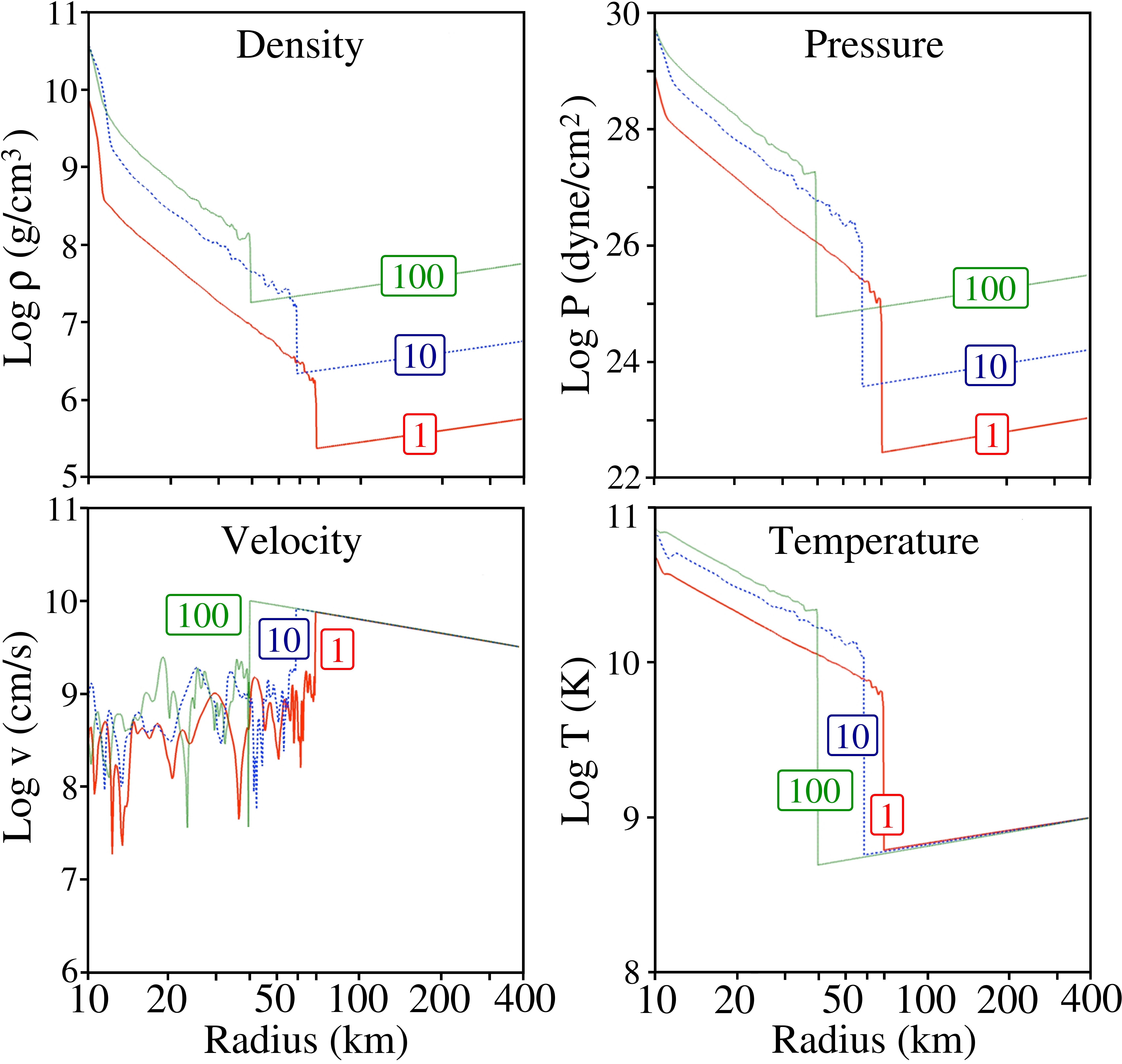}
\end{center}
\caption{
Radial profiles of density, pressure, velocity, and temperature, for highly hypercritical accretion rates 
$\dot{m} = (1, 10, 100) \cdot \dot{m}_0$ (labeled), in the 2D free-fall case,
once the stationary state has been reached.}
\label{Fig:3}
\end{figure}

\begin{figure*}
\begin{center}
\includegraphics[width=0.9\textwidth]{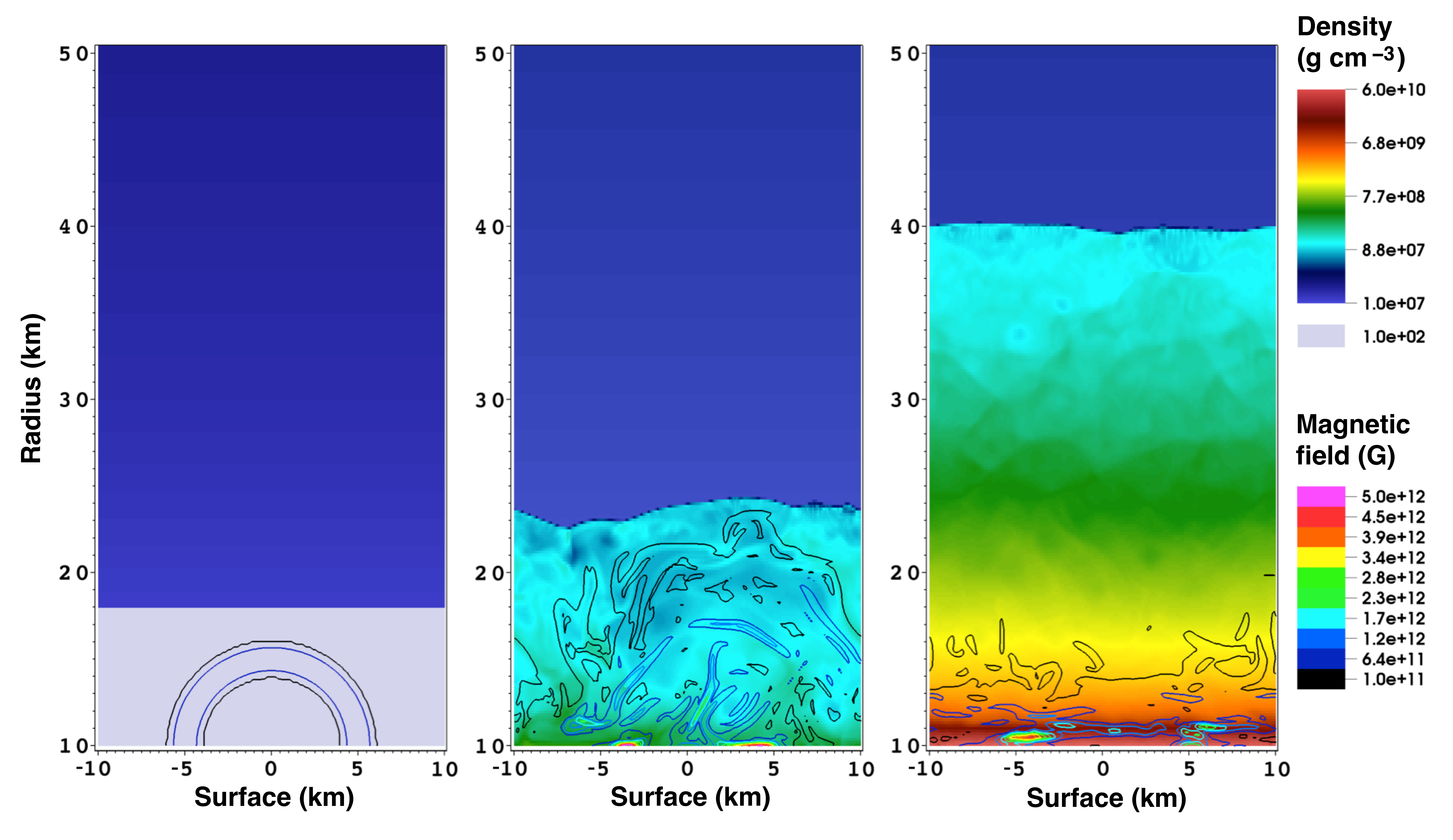}
\end{center}
\caption{
Density maps (color) with magnetic field contours superimposed,
at three instants, $t=0$, 1, and 100~ms,
for a 2D MHD simulation in the free-fall scenario, with $\dot{m}=100\, \dot{m}_{0}$. 
The imposed level of refinement was 4, with 3 blocks in $x$-direction
and 39 in the $y$-direction, which results in $192 \times 2496$
effective zones in the computational domain. Note how the shock rises progressively 
and the magnetic field is submerged close to the neutron star surface. 
The anchor points of the magnetic loop are still visible in the final snapshot.
}
\label{Fig:4a}
\end{figure*}

\begin{figure*}
\begin{center}
\includegraphics[width=0.9\textwidth]{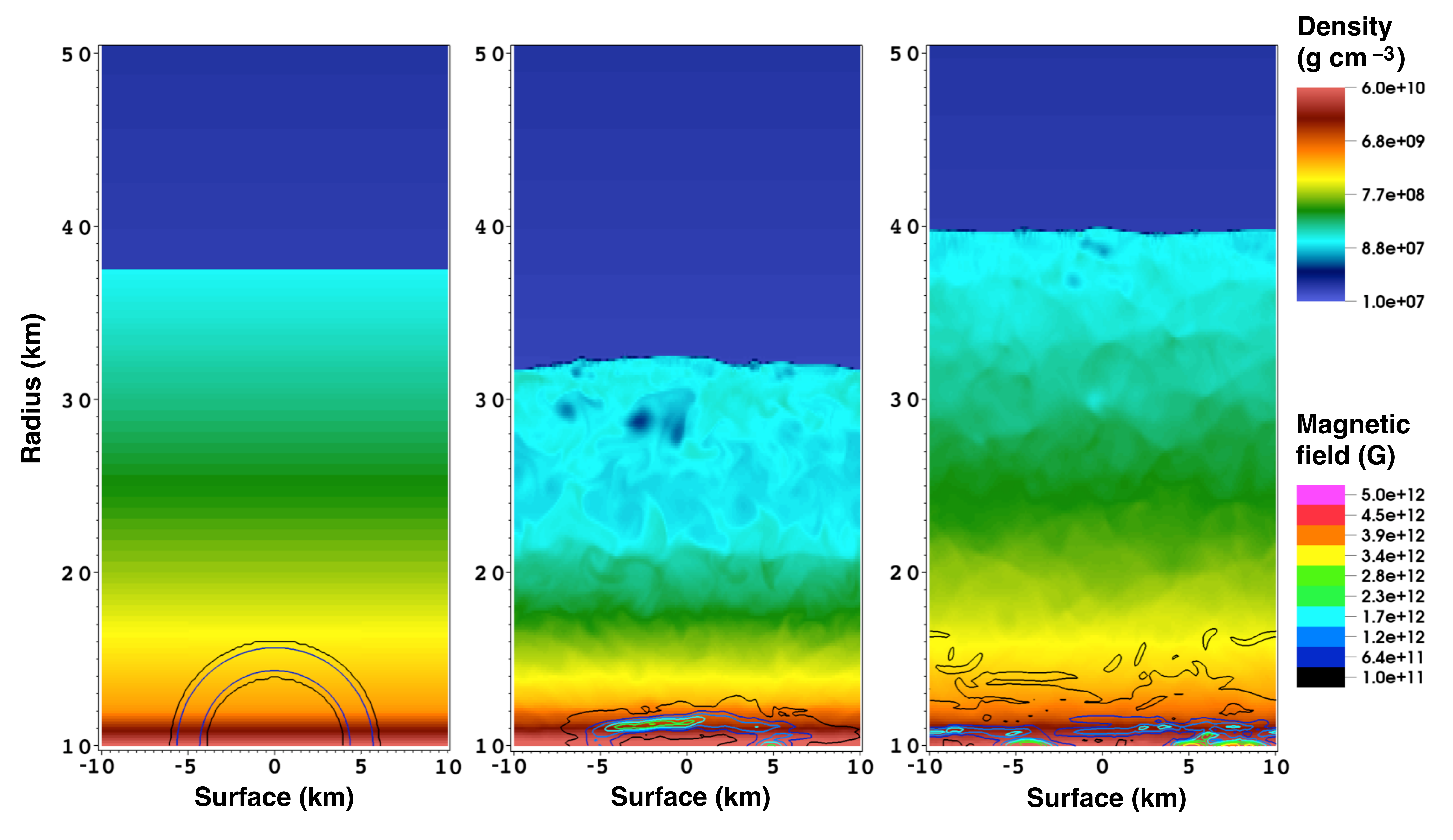}
\end{center}
\caption{
The same as in Figure~\ref{Fig:4a}, at the same times $t=0$, 1, and 100~ms,
but for the initial atmosphere in quasi-hydrostatic equilibrium (QHEE). 
After the initial transient, the flow settles also to a quasi-static atmosphere, 
with the magnetic field being strongly confined to the vicinity of the neutron star surface. 
The anchor points of the magnetic loop are also visible in the final snapshot.
}
\label{Fig:4b}
\end{figure*}

We describe here results for highly hypercritical accretion rates,
$\dot{m} = (1, 10, 100) \cdot \dot{m}_0$, for which we are able to follow the evolution of the accretion shock. 
In Figure \ref{Fig:3} we plot their final radial profiles of density, pressure, velocity and 
temperature for 2D simulations in the free-fall scenario. 
Here the final state is defined once the accretion shock, and the included envelope, 
have reached a quasi-stationary state.
In the density and pressure profiles the piling up of matter close to the neutron star surface is notorious 
(an effect not accounted for in the analytical approach).
Notice that, although the system has reached a  quasi-stationary state in all cases,
significant noise remains in the velocity profile: 
because of the periodic boundary condition on the vertical sides
matter can freely flow in the horizontal direction, as well as bounce off the neutron star surface,
which prevents a full stationary state from being reached. 
Nevertheless, the mean velocity profile is close to the analytical one, 
indicating that the system has largely relaxed despite these fluctuations.
In these free-fall simulations, for our fiducial accretion rate $\dot{m}_{0}$ the system reaches the 
quasi-stationary state in about 600 ms, whereas for higher accretion rates this time is substantially reduced:
300 ms for $10\, \dot{m}_{0}$ and 100 ms for $100\, \dot{m}_{0}$. 

We show in Figure~\ref{Fig:4a}, for a $100\, \dot{m}_{0}$ accretion rate,
color maps of density with magnetic field contours superimposed, for the 2D free-fall scenario.
The initial panel at $t=0$ shows the reverse shock just before it reaches the magnetic field loop.
When the reverse shock hits the NS surface and bounces a violent, convective,
layer appears, the accretion shock develops and rapidly moves upward against the infalling matter
($t=1$ ms panel).
The flow smoothens gradually as the accretion shock stabilizes.
Instabilities of the Rayleigh-Taylor type are present in this regime, but they
disappear when the system reaches equilibrium (see the panel at $t=100$~ms).
The magnetic loop is immediately torn apart by the reverse shock
and the initial strong convection which is dragging the field with it within 
the developing envelope (panel at $t=1$~ms).
Although it has a turbulent dynamics,
the magnetic field, having its feet frozen into the neutron star surface, 
always remains confined below the accretion shock.
After 50~ms, the magnetic field begins to be submerged
and is trapped into the material that is piling up onto the neutron star surface. 
The matter outside the accretion shock is still falling with constant accretion rate, 
but it is the fluid inside the envelope, with its much lower downward velocity, which is nevertheless
responsible for the magnetic field submergence. 
After 100~ms, the submergence is completed, with a maximum magnetic field strength 
$\simeq 5 \times10^{12}$ G. 
At $t=0$ the magnetic loop was immersed in a medium of density $10^2$ g cm$^{-3}$,
the infalling matter just behind the reverse shock had a density $\sim 10^7$ g cm$^{-3}$
but after 100~ms the magnetic field has been submerged into matter at density 
$\simgreater 5 \times 10^9$ g cm$^{-3}$.

For comparison, in Figure~\ref{Fig:4b} we show the results of a similar simulation but
within the QHEE scenario.
At $t=0$ the magnetic field loop is immersed within the QHEE.
The initial, strongly convective phase seen in the free-fall scenario is absent.
Some weak convection is nevertheless present in the upper part of the envelope,
whose height initially oscillates ($t=1$ panel in Figure~\ref{Fig:4b}).
Due to the smoother evolution, compared to the free-fall scenario,
even at this early time the magnetic loop has already been compressed by the infalling
envelope but is not strongly disrupted as in the free-fall scenario.
After 100 ms, the QHEE has recovered its initial shape and height,
with an equilibrium shock radius $r_{sh}\simeq 4 \times10^6$ cm,
and the magnetic field is seen to be submerged.
Although the accretion velocity within the envelope is much lower than above the shock,
it is enough to completely submerge the magnetic loop.

In the free-fall scenario the initial smashing of the magnetic loop by the reverse shock
is immediately counteracted by the bouncing of the matter off the neutron star surface
which drags the field with it, upward, within the forming envelope.
The field submergence in the QHEE scenario at early times shows that it is the slow settling of the
envelope material, under continuous accretion, which is actually responsible for the field submergence.
In the free-fall scenario the field submergence follows the same route as in the QHEE scenario
once the envelope stabilizes and its matter slowly settles onto the neutron star surface.
The final geometry of the magnetic field is seen to be very similar under both free-fall and QHEE
initial conditions.

\begin{figure}
\begin{center}
\includegraphics[width=0.4\textwidth]{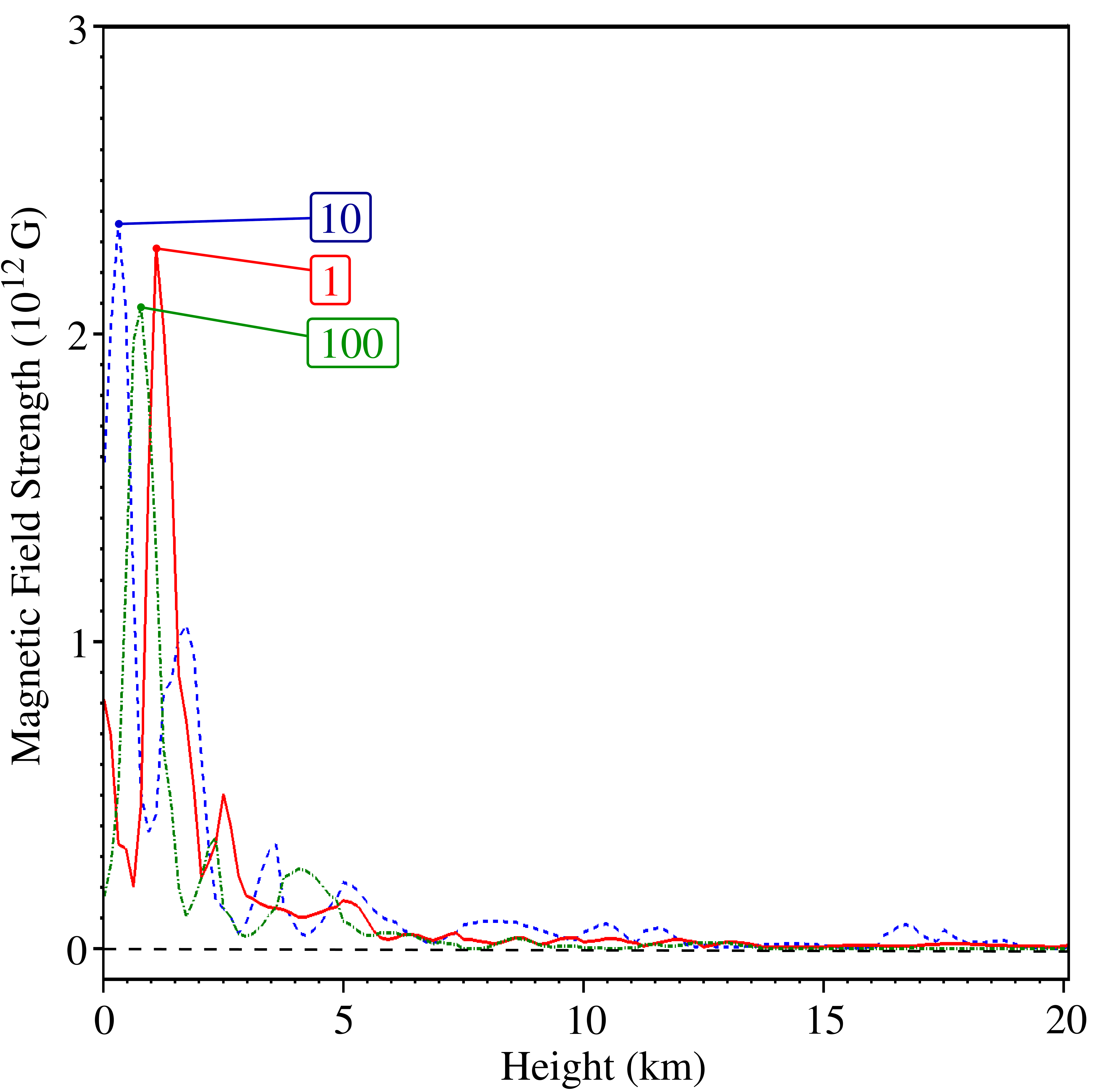}
\end{center}
\caption{
Radial profiles of the magnetic field strength, once the stationary state has been reached,
in the free-fall scenario,
for $\dot{m} = \dot{m}_{0}$ at time $t = 600$ ms, 
$\dot{m} = 10 \, \dot{m}_{0}$ at time $t = 300$ ms, and
$\dot{m} = 100 \, \dot{m}_{0}$ at time $t = 100$ ms.
The horizontal averages of the field strength as a function of the
height above the neutron star surface are plotted for each case (labeled).
}
\label{Fig:5b}
\end{figure}

For the lower accretion rates, 10 and $1 \, \dot{m}_{0}$, the evolution is very similar but
the accretion shock reaches much larger heights.
We choose to illustrate only the  $100\, \dot{m}_{0}$, in Figures~\ref{Fig:4a} and \ref{Fig:4b}
since the lower rate cases involve much higher columns.
We show in Figure~\ref{Fig:5b} the radial profiles of the magnetic field for
our three highly hypercritical accretion rates, in the free-fall scenario, corresponding to the times
at which the quasi-hydrostatic equilibrium has been reached.
In the three cases the submergence of the magnetic field is clearly seen.

It is thus clear from the above results that for strongly hypercritical mass accretion rates, 
both approaches, the free fall and QHEE initial conditions, lead to the same conclusion, 
namely, that the field is promptly submerged by the sheer force of the
settling fluid in the envelope. 
Thus we believe this to be a robust result reflecting the actual fate likely to 
befall an initial magnetic field anchored on the neutron star
once a quasi-stationary settling envelope has been established.

\subsubsection{3D simulation}
\label{Sec:3D}

In all the high accretion rates described above we observe the submergence of the
magnetic field.
These were, however, 2D geometries and, since the magnetic field behavior
is an intrinsically 3D phenomenon, one may wonder how much the addition of the
third dimension degree of freedom would alter the results.
Moreover, in a 3D environment the  accreting matter has the possibility to slip around the
magnetic field loop, hence potentially reducing the pressure it exerts on it.
To tackle this issue, we performed a 3D MHD simulation for the free-fall scenario.
From the available computing power we were limited to simulate only a short
column, hence we consider only our highest accretion rate, $100 \, \dot{m}_{0}$.

\begin{figure*}
\begin{center}
\includegraphics[width=0.80\textwidth]{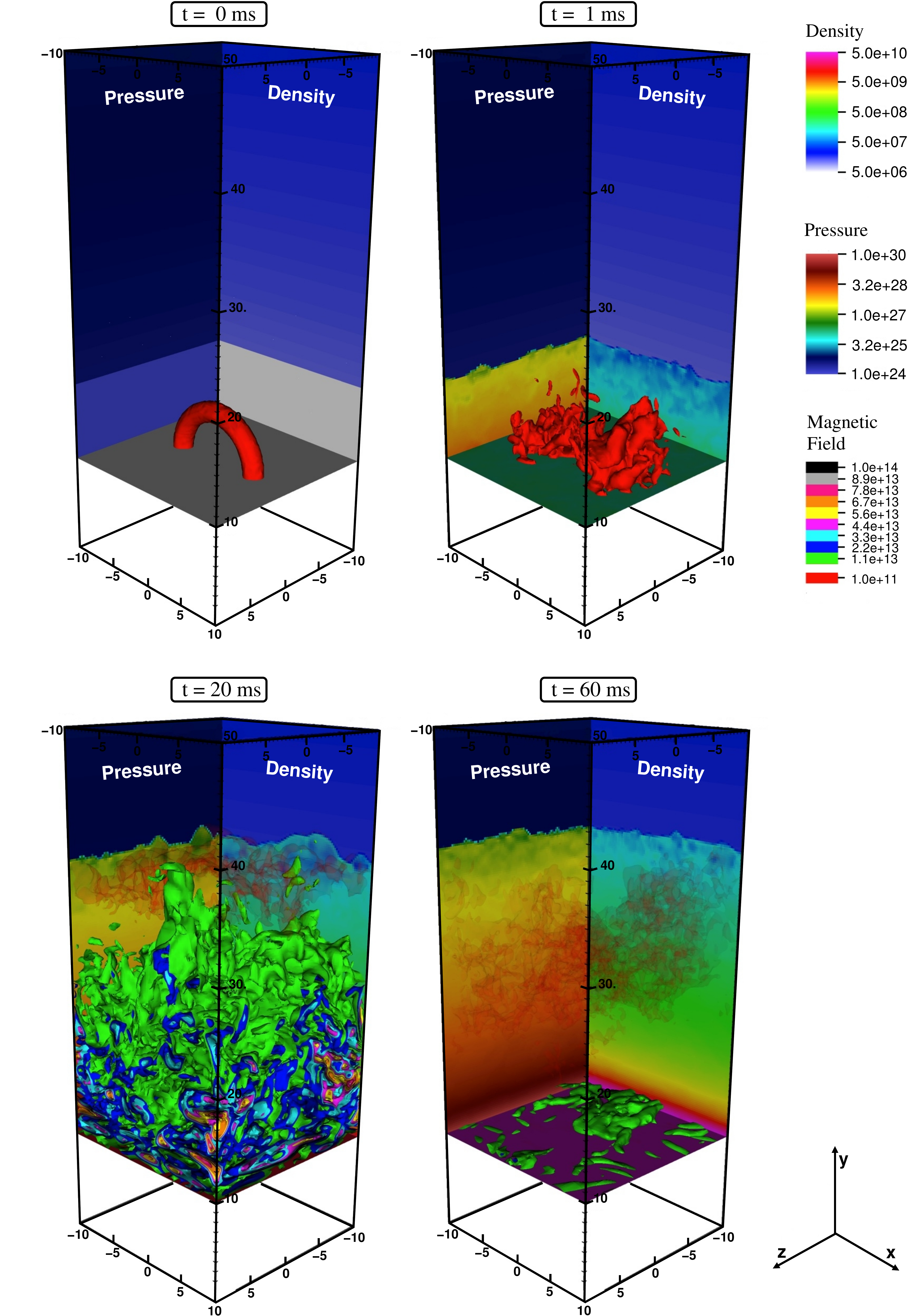}
\end{center}
\caption{
Iso-surfaces of magnetic field at times $t$=0, 1, 20, and 60~ms,
for the 3D simulation in the  free-fall scenario, and an accretion rate $\dot{m}=100 \, \dot{m}_{0}$.
We show color maps of density and pressure in the $xy$ and $yz$ faces, respectively.
In the two upper panels only the $10^{11}$ G iso-surfaces are visible, the magnetic field
being, however, stronger within these surfaces.
In the two lower panels the $10^{11}$ G iso-surface is rendered with high transparency
allowing to clearly see much stronger magnetic field regions. 
The end result of magnetic field submergence  due to the strong accrretion is the same as in the 2D simulations.
} 
\label{Fig:6}
\end{figure*}

In Figure~\ref{Fig:6} we show four snapshots of the time evolution of the magnetic field
iso-surfaces until the quasi-statonary state is reached. 
Notice that the fluid reaches the stationary
state in 60 ms, i.e. in a shorter time than in the corresponding 2D simulation.
Overall, the evolution is very similar to the corresponding 2D simulations.
After 1 ms the magnetic field loop has been smashed by the accretion shock
and the field is beginning to be spread by the violent fluid motion.
The accretion shock rapidly progresses upward and reaches its final height,
$r_{sh} \simeq 4 \times 10^6$ cm as in the 2D case, in about 20 ms.
Notice, at that time, that the magnetic field is also being dragged upward by the
fluid motion but is not able to reach the location of the shock.
In no moment will the field be able to rise above the forming envelope.
However, when the QHEE has built up and the initial strongly convective regime has died away
the magnetic field has been forced downward by the slow settling of matter in the envelope
and is submerged inside the denser layers onto the neutron star surface.
Notice that the magnetic field has been spread all over the whole
bottom area of the simulation box.

\begin{figure}
\begin{center}
\includegraphics[width=0.45\textwidth]{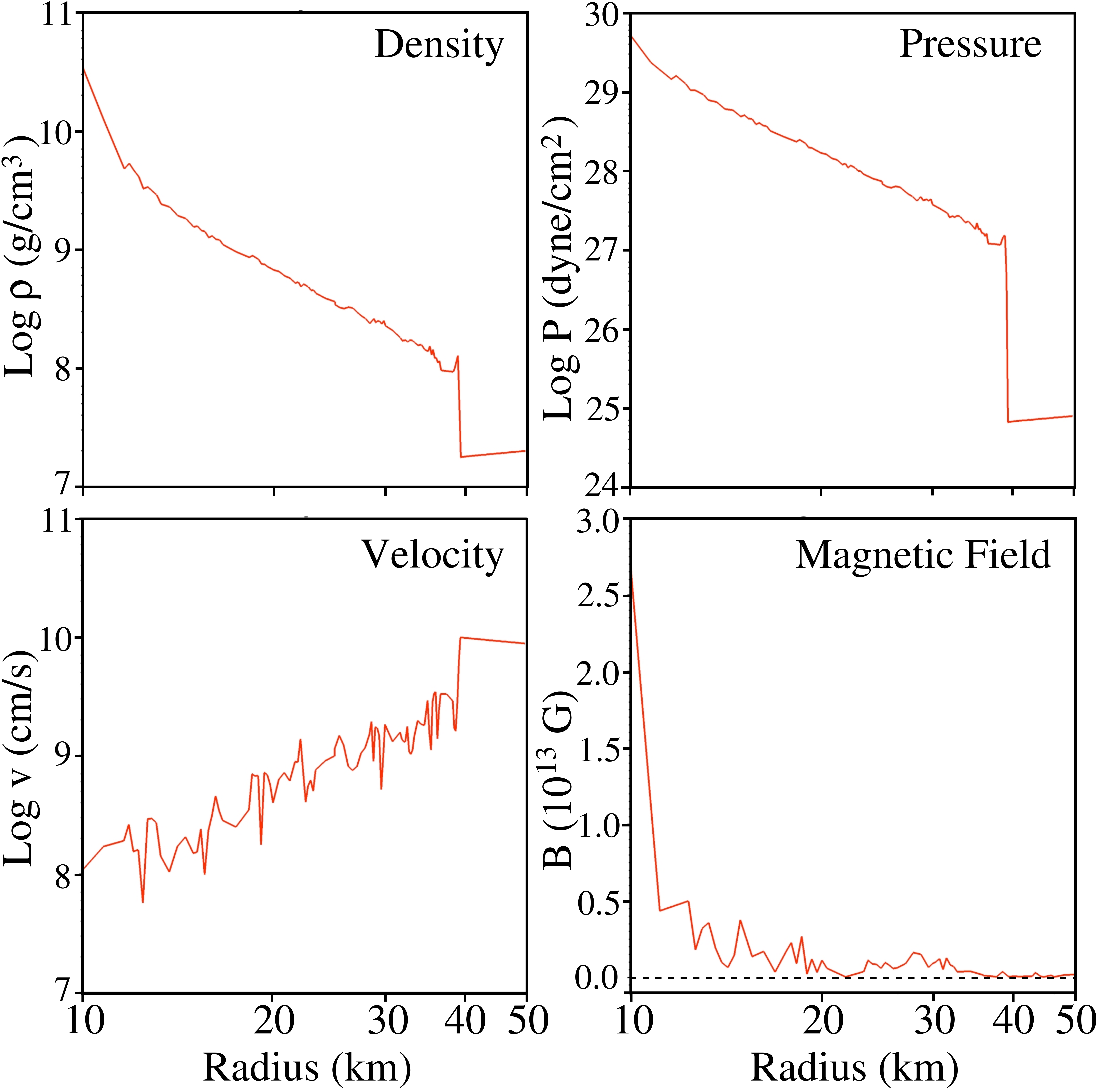}
\end{center}
\caption{
Radial profiles of averaged density, pressure, velocity and magnetic field
for the 3D model depicted in Figure~\ref{Fig:6},
once the stationary state has been reached, 
at $t = 60$~ms. Comparison with Figure~\ref{Fig:3} shows that the solution in terms of 
position of the shock front and values is the same as in 2D. 
}
\label{Fig:7}
\end{figure}

In Figure~\ref{Fig:7} we show the radial profiles of the density, pressure,
velocity and magnetic field in the last stage. 
The density and pressure profiles are essentially identical to the 2D ones shown in
Figure~\ref{Fig:3} with the clear piling up of matter onto the neutron star surface.
The velocity profile, however, is much less noisy in 3D than in 2D.
These three, $\rho$, $P$, and $v$, profiles are also in close agreement with the analytical solution.
The magnetic field strength profile clearly exhibits the submergence within the high density region. 
We thus conclude that at this level of mass accretion, no significant differences in morphology 
are apparent by considering the 3D nature of the problem, and our previous statements 
concerning the submergence of the field from the 2D calculations remain valid. 

As a further energy check, the neutrino luminosity integrated over the whole domain,
at $t = 60$ ms when the quasi-stationnary state has been reached, is 
$L_\nu \simeq 2.3 \times10^{50}$ erg s$^{-1}$
but the emission is strongly concentrated within the first two kms above the stellar surface
due to the strong temperature dependence of the emissivity.
The gravitational energy liberated by the accretion is $\sim 0.3 \times \dot{m} c^2
\simeq 2.1 \times 10^{50}$ erg s$^{-1}$ and is hence emitted essentially in its entirety in neutrinos.

Once the quasi-stationary state has been achieved, at $t = 60$ ms, 
the adiabatic indices connected with the sound
velocity and with the system energy are: $\gamma_{c}=1.35$ and $\gamma_{e}=1.34$.
Then the adiabatic and radiative gradients are: $\nabla_{ad}=1-1/\gamma_{c} \simeq 0.26$,
$\nabla_{rad} = (d\ln T/d\ln P) \simeq 0.25$. 
The system is, hence, convectively stable, but only marginally.

\subsection{Weakly hypercritical accretion rates}
\label{Sec:lowrates}

\begin{figure*}{b}
\begin{center}
\includegraphics[width=0.95\textwidth]{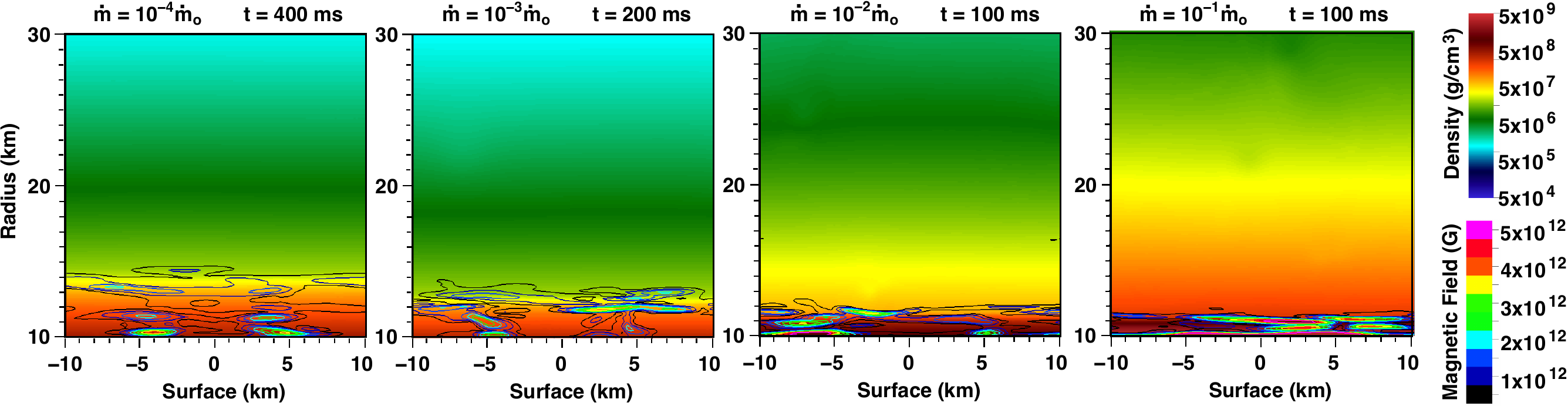}
\end{center}
\caption{Color maps of the density with magnetic field intensity contours superimposed,
for the four weakly hypercritical accretion rates in the initially QHEE scenario:
$10^{-1}$ and $10^{-2} \, \dot{m}_0$ at time $t = 100$ ms,
$10^{-3} \, \dot{m}_0$ at time $t = 200$ ms, and
$10^{-4} \, \dot{m}_0$ at time $t = 400$ ms. 
The submergence of the field is now clearly affected by the assumed accretion rate. 
 }
\label{Fig:8}
\end{figure*}

We can now investigate what happens with the magnetic loop anchored onto the
neutron star surface for lower accretion rates.
To answer this, we carry out 2D simulations for the cases 
$\dot{m} = (10^{-4}, 10^{-3}, 10^{-2},10^{-1})~\cdot~\dot{m}_0$.
For these lower accretion rates it is computationally impossible to follow the expansion
of the accretion shock because it reaches to extremely large heights, 
leaving the computational domain, and, moreover, the relaxation time of the envelope becomes exceedingly long.
Attempting to follow the evolution of a free-fall scenario by allowing the accretion shock to leave
the integration domain would be physically, and numerically, inconsistent with our setting:
it not possible to impose an upper boundary condition which incorporates in any reasonable
manner the strongly turbulent motion of the initial phase of envelope formation.
Due to this, we are forced to restrict ourselves to the QHEE scenarios.
The results of our study of the highly hypercritical accretion rates in \S~\ref{Sec:highrates}
showed that the field submergence is occurring in a very similar manner in both types of scenarios
and that, in the long term, it is manly due to the settling of matter onto the neutron star surface.
This gives us confidence that limiting ourselves to the QHEE scenario for the weakly hypercritical 
rates will give us representative results.

\begin{figure*}
\begin{center}
\includegraphics[width=0.6\textwidth]{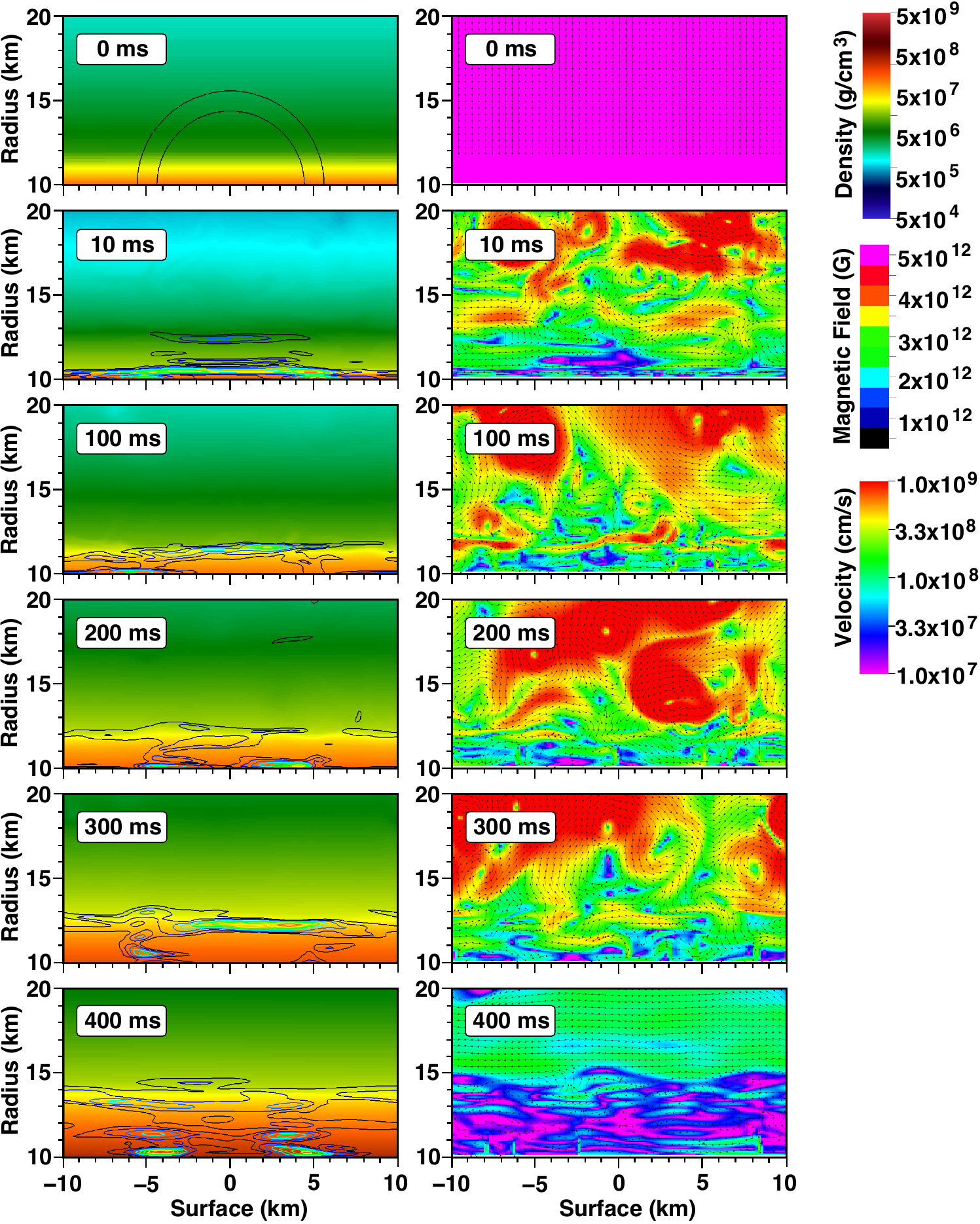}
\end{center}
\caption{Evolution of the accretion flow in the QHEE initial condition scenario with 
$\dot{m} = 10^{-4} \, \dot{m}_0$ at 6 different times, as labeled.
Left panels: color maps of the density with magnetic field intensity contours superimposed.
Right panels: color maps of the velocity with arrows showing its direction. 
After an initialy transient in which the field is rapidly compressed, it rises to a height of $\simeq 4$~km
(radius $\sim 14$ km).
The change in the pattern of the flow across the boundary from low (at top) to high (bottom) density is apparent.
}
\label{Fig:9}
\end{figure*}

\begin{figure}
\begin{center}
\includegraphics[width=0.35\textwidth]{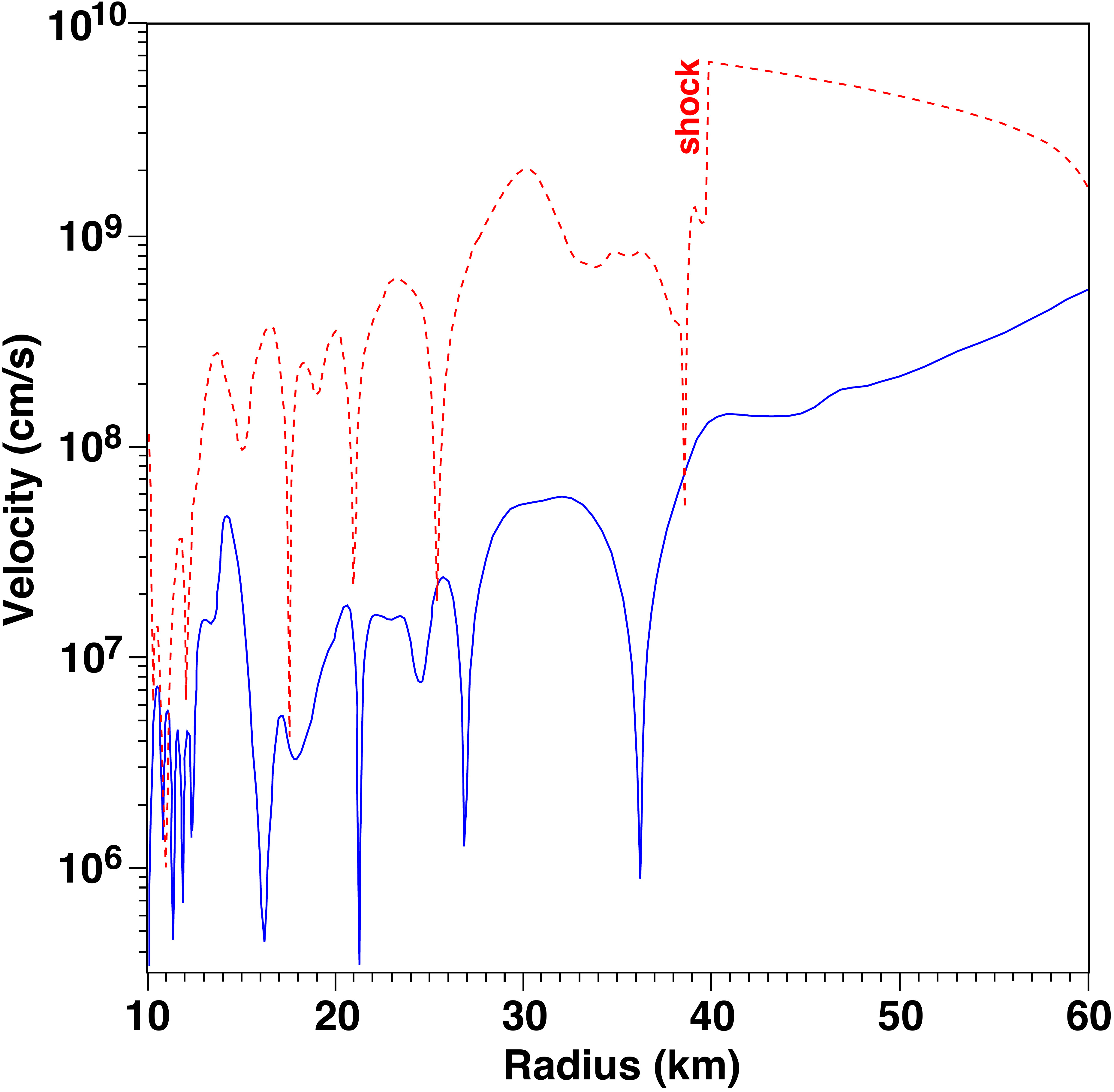}
\end{center}
\caption{Profiles of the absolute value of the vertical velocity, $|v_y|$,
in the $\dot{m} = 10^{-4} \, \dot{m}_0$ scenario at $t=10$~ms (dotted curve) and $t=400$~ms (solid curve). 
The position of the shock from the infalling matter at 10~ms is labeled. 
The regular  fluctuations in velocity are due to the convective motions in the flow.
The size of the cells can be readily measured and the widely different lengh-scales
between the high density forming crust and the envelope are clearly seen.
}
\label{Fig:10}
\end{figure}

We simulated an accretion column with height $\Delta y=4\times10^6$ cm,
using the upper boundary condition with matter injection at the same $\dot{m}$
and the corresponding $\rho$ and $v$ as given by the analytical solution for a QHEE.
The initial magnetic field configuration was that of a magnetic loop,
as described in \S~\ref{Sec:Model}, immersed within a QHEE envelope 
Each simulation exhibited a short initial relaxation phase but the system rapidly reached stationary 
density and pressure profiles.
In Figure~\ref{Fig:8} we show color maps of density, with magnetic field contours
superimposed, for these four accretion rates at times where a stationary state was well established
in the density profile.
At the highest accretion rate, $\dot{m} = 0.1 \, \dot{m}_0$, in 100 ms the magnetic loops is clearly
submerged within the high density bottom of the envelope.
This result is similar to what we obtained at the highly hypercritical rates in \S~\ref{Sec:highrates}.

At lower rates, $\dot{m} \ll 0.1 \, \dot{m}_0$, however, the situation becomes more complex.
As seen in Figure~\ref{Fig:8}, the magnetic field is mostly confined in the high density region
but is able to extend to slightly larger heights than before.
The evolution of our lowest accretion rate scenario, $\dot{m} = 10^{-4} \, \dot{m}_0$, is depicted in
greater detail in Figure~\ref{Fig:9}, a simulation we could pursue for 400 ms
(after that time the turbulence reached the upper boundary of our integration domain
and conflicted with the inflow condition, so the simulation was stopped).
The left panels show that, after an initial strong compression, the field is slowly rising:
at 100~ms it is essentially confined within the high density region but, as matter piles up onto the neutron
star surface and this high density region expands upward, the magnetic field follows it.
The right column in Figure~\ref{Fig:9} illustrates the pattern of convective cells with very high velocities in the
upper low density regions that contrast with the much slower motion within the much denser matter 
accumulating onto the neutron star surface.
The size of the convective cells is expected to be of the order of the pressure scale-height, 
$H_P = -dr/d\log P$, which is $\sim$ 10~km in the envelope but only $\sim$ 1~km in the forming crust.
These two distinct patterns, and their length-scales, are clearly seen in Figure~\ref{Fig:10}, 
where the absolute value of the vertical velocity is plotted as a function of radius,
and are conserved during the evolution in spite of a more than one order of magnitude decrease in velocity.
It is this clear difference in the two convection patterns, large, low density cells with high velocites 
on top of small, high density cells with low velocities
that keeps the crushed magnetic field confined into the high density region of the forming new crust.

\section{Discussion and Interpretation}
\label{Sec:Discussion}

We have continued our study of hypercritical accretion onto newborn neutron stars,
in particular with the aim of studying the possible submergence of the magnetic field for various accretion rates. 
This is particularly relevant in the context of making the neutron star eventually invisible as a pulsar 
following the supernova explosion. 
Extending our exploration of parameter space to wide columns spanning a significant fraction of the stellar surface, 
we have paid special attention to the formation of the quasi-steady atmosphere that forms after the reverse shock 
has reached the hard surface of the star, and how it affects an initial loop of field anchored in it.
We find that whether an initial condition in free fall, or an analytical atmosphere in quasi-hydrostatic equilibrium, is initially used, 
the field is rapidly submerged into the forming high-density new crust for strongly hypercritical rates, 
above our fiducial value of $\dot{M}=\dot{M}_{0}=350$~M$_{\odot}$~yr$^{-1}$. 
The comparison between results for  different initial conditions is important because it is extremely challenging 
to follow the evolution of the system, in particular the shock front, for lower accretion rates. 
This is due to the scaling of the equilibrium position of the shock, given in Equation \ref{eq:ysh},
that shows the shock expands to very large heights at small accretion rates.
We have further carried out a calculation in three dimensions, to investigate whether the flow exhibits qualitative changes, 
in particular regarding the submergence of the field, as compared to the 2D simulations. 
Finding none, we conclude that 2D studies are able to correctly capture the main features of the 
field's behavior in this respect.  

As the accretion rate is lowered, reaching down to 10$^{-4}$~$\dot{M}_0 \sim 0.035 M_\odot$ yr$^{-1}$, 
the field is progressively less affected by the infall, and the initial loop reaches upwards of the neutron star surface 
after the initial transient, to a height of a few kilometers.
Intringuingly, the pattern of convective cells helps explain why the field is not dragged to greater heights.
Close to the surface of the star, a high-density region where matter accumulates is present
and this is also where most of the magnetic field remains confined.
As described in \S\ref{Sec:lowrates}, there is a sharp difference in the convection patterns in the high density region,
with cell sizes $\sim 1$ km, compared to the overlaying envelope, with cell sizes $\sim 10$ km.
Thus, the envelope plasma circulations are unable to reach into the region of high field and drag it to larger altitudes.
Notice that even at the lowest accretion rate we considered, the matter pressure is much larger than the magnetic 
pressure and the field is continuously dragged by the motion of the plasma.

Our distinction between highly and weakly hypercritical accretion rates is purely due to our computational limitations.
However, we do find a {\em physical} distinction in the field submergence that is simply due to the evolution of
the convection in the envelope and the forming crust.
In the highly hypercritical cases of \S\ref{Sec:highrates} and the highest rate, $0.1 \dot{m}_0$, of \S\ref{Sec:lowrates}
our simulation lasted long enough for convection in the envelope to essentially disapear,
while at the lower rates significant convection was always present.
In the former cases the only remaining fluid motion is the slow settling of the accreting matter onto the neutron star
and, since the magnetic field is frozen into the plasma, the field is completely submerged into the high density region
of the forming crust.
This result is still clearly seen in the $0.1 \dot{m}_0$ case of Figure~\ref{Fig:8}.
In contradistinction, in the latter cases continued convection maintains mixing of the magnetic field, but the different patterns of
convection in the forming crust compared to the envelope traps most of the magnetic field in the high density
regions and the field present in the envelope is strongly reduced compared to its initial value.
The evolution displayed in the left panels of Figure~\ref{Fig:9} clearly shows the slow submergence of the field 
in the high density region as matter piles up onto the neutron star and the continuous presence of a residual
field in the bottom of the envelope in the interface region between the two regions of different convection patterns.
Notice that the apparent rise of the magnetic field in this Figure~\ref{Fig:9} is actually just the rise of the height of the
building new crust while the residual field never extends into the envelope much beyond the interface region.
If this accretion phase last for several hours, most of the magnetic field should eventually be pushed deep into the crust.

{\em The critical factor separating a total versus a partial submergence of the initial magnetic field hence seems to be the time 
scale for convection in the accretion envelope to disappear.}
Once convection stops the magnetic field is unavoidably submerged into the neutron star curst. 
As described by \cite{Fryer:1996qf}, at low accretion rate the development of a stationary non-convective envelope may take more
time than the duration of the hypercritical accretion phase.
Supernovae in which a strong reverse shock was produced and in which the accreting envelope reached a convectively stable
profile are likely to produce a neutron star with a vanishingly small surface magnetic field.
In contradistinction, supernovae that produced a weak reverse shock whose accreting envelope remained convective
can be expected to produce weakly magnetized neutron stars. 
In the case where no reverse shock developed, the neutron star will simply maintain the magnetic field it had at its birth.

\section{Conclusions}
\label{Sec:Conclusions}

We have extended our previous study of Paper I of the effect of post-supernova hypercritical 
accretion on the magnetic field of the new-born neutron star.
Our study is still restricted to cases where the accreting matter pressure dominates over the magnetic pressure,
so that the magnetic field is only a ``victim'' of the accretion.
We found that the critical factor determining the fate of the magnetic field is whether the developing
quasi-hydrostatic equilibrium accreting envelope will reach of state of convective stability or not.
Once convection stops the magnetic field is unavoidably pushed into the forming new crust and
a non magnetized neutron star will result. 
If the acreting envelope remains convective, most of the initial magnetic field is still pushed into the
forming crust but the fluid motions, due to the diferent patterns of convection in the crust and the
surrounding envelope, maintain a residual field protruding out of the star.
In this latter case, the resulting external field certainly has a very complicated structure
and anisotropy in the accretion, as, e.g., from rotation, will also increase the compexity of the final field.
The dipolar component of the final field may be several orders of magnitude smaller than the initial one
but localized spots with a much stronger field seem unavoidable.

Non-convective envelopes are rapidly obtained only at extremely high accretion rates,
larger that $\sim 10 M_\odot$ yr$^{-1}$ in our study. 
It is not clear how many new-born neutron stars undergo such events, and how many of them would survive
without collapsing into a black-hole \citep{Fryer:1996qf}.
Our conclusion is that there may be only very few young neutron stars that have been totally de-magnetized
by hypercritical accretion and they are likely very massive ones.

At lower accretion rates, field submergence is only partial and the fraction of new-born neutron stars
having passed through such events is likely significant. 
\cite{Fryer:1996qf}  were very careful in the treatment of neutrino heating and found that when 
$\dot{M} > 0.1 M_\odot$ yr$^{-1}$ a neutrino driven explosion may result, expelling the envelope
while at lower rates steady accretion is achieved with, eventually, a non-convective envelope.
As the accretion rate decreases one may reach a point where the residual magnetic field may begin to play
a dynamical role.
The final outcome of such evolution is impossible to treat numerically and can only be guessed at now,
but a weakly magnetized neutron star, with a complicated field geometry seems very likely.

The CCOs are the primary candidates for neutron stars having undergone such a field submergence
and they constitute half of the known population of young neutron stars:
\cite{Popov:2012fk} list a dozen of pulsars with age $< 10$ kyrs, three of them being CCOs,
i.e. a total of nine non-CCOs pulsars,
while \cite{Gotthelf:2008fu} present eight CCOs, all having ages below 10 kyrs
to which one may add the yet undiscovered neutron star in the remnant of SN 1987A.
The three CCOs with measurements of spin periods $P$, and also derivatives $\dot P$
from which surface dipolar field strengths $B_s$ can be inferred,  are 
PSR J1852+0040 in Kesteven 79, with a period $P=105$ ms and $B_s = 3.1\times 10^{10}$~G \citep{Halpern:2007uq},
and 
PSR J1210-5226 in PKS 1209-51/52, with $P=424$ ms and $B_s = 9.8 \times 10^{10}$~G
and 
PSR J0821-4300 in Puppis A with $P=112$ ms and $B_s = 2.9 \times 10^{10}$~G
(Gotthelf \& Halpern to be published in ApJ).

The unavoidable prediction of a submerged magnetic field is the subsequent growth of the surface
field by back-diffusion of the internal field \citep{Muslimov:1995nx}.
Hence, burial of the field in a newborn neutron star may manifest itself as a delay in the star turning 
on as a classical radio pulsar.
Possible evolution scenarios were studied in 1D models in \cite{Geppert:1999cr} and \cite{Ho:2011pi}
and recently by \cite{2012MNRAS.425.2487V} in a 2D scenario.
A growing magnetic field also naturally manifests itself through a braking index inferior to 3
\citep{Muslimov:1996vn,Muslimov:1999ys} and may lead to peculiar evolutionary properties, 
such as those already observed in PSR J1734-3333 \citep{Espinoza:2011fk}. 

In the present work we were still dealing with magnetic field strengths, and accretion rates, in which
the magnetic forces are negligible, even though the FLASH code does take them into account.
More interesting effects can be expected in the case of a magnetar size field \citep{Thompson:2001zr}
and will be considered in a future paper.

Finally, among the many simplifications of our model, the assumption of non magnetized accreted matter 
is clearly disputable.
In case the neutron star magnetic field is of fossile origin, the accreting flow is likely bringing in some
significant magnetic field that may just replace the submerged one.
One may conceive that accretion will hence complicate the geometry of the surface field without necessarily
reducing its strength, but possibly reducing the strength of the dipolar component.
However, turbulence in the accretion envelope can act as a diamagnetic medium \citep{Vain:1972ab}
which may also result in a strong suppression of the surface magnetic field.
Such an evolution certainly deserves further study.

\acknowledgements 

We are grateful to DGTIC-UNAM for allowing us to use its
KanBalam Cluster, where all the simulations were performed. 
The software used in this work was in part developed by the 
DOE NNSA-ASC OASCR Flash Center at the University of chicago
This work was supported in part by CONACyT grants 
CB-2009-1 \#132400
and CB-2008-1 \#101958.



\end{document}